\documentclass[11pt,a4paper,english,superscriptaddress,aps,preprint,showpacs]{revtex4}
\usepackage{amsfonts}
\usepackage{graphics}
\usepackage{graphicx}
\usepackage{slashed}
\usepackage{amsmath}
\usepackage{amssymb}
\makeatletter
\usepackage{babel}
\newcommand{\bea}{\begin{eqnarray}}
\newcommand{\eea}{\end{eqnarray}}

\begin{document}

\title{On the one-loop effective potential in nonlocal supersymmetric theories}

\author{E. R. Bezerra de Mello, F. S. Gama, J. R. Nascimento, A. Yu. Petrov}

\affiliation{Departamento de F\'{\i}sica, Universidade Federal da 
Para\'{\i}ba\\
 Caixa Postal 5008, 58051-970, Jo\~ao Pessoa, Para\'{\i}ba, Brazil}
\email{emello@fisica.ufpb.br,fisicofabricio@yahoo.com.br,jroberto@fisica.ufpb.br,petrov@fisica.ufpb.br}

\begin{abstract}
Within the superfield approach, we consider the nonlocal generalization of the Wess-Zumino model and calculate the one-loop low-energy contributions to the effective action. Four different nonlocal models are considered, among which only the first model does not reduce to the standard Wess-Zumino model when we take the parameter of nonlocality of the model, $\Lambda$, much greater than any energy scale; in addition, this model also depends on an extra parameter, $\xi$. As to the other three models, the result looks like the renormalized effective potential for the usual Wess-Zumino model, where the normalization scale $\mu$ is replaced by the $\Lambda$. Moreover, the fourth model displays a divergence which can be eliminated through the appropriate wave function renormalization. 
\end{abstract}
\pacs{11.30.Pb, 11.10.Ef}

\maketitle

\section{Introduction}

The one-loop  K\"ahlerian effective  potential in a general $N=1$ supersymmetric quantum field theory has been  calculated in \cite{Grisaru96} many years ago.  In that paper, a regularized result is obtained by introducing an ultraviolet cutoff energy scale. A possible way to overcome the divergence problem can be implemented by considering a new class of theories, that is, the nonlocal ones, which,  from one side, preserve Lorentz symmetry in a manifest form, from another side, allow to avoid the arising of ghosts (within the gravity context, such a possibility has been discussed in \cite{Modesto,Biswas}). In this way a nonlocal extensions of a scalar quantum field theory have been recently considered in \cite{Briscese15}. Therefore, the natural question would consist in the generalization of this approach for the supersymmetric field theory, especially within the superfield formalism which is know to be the most convenient description of the supersymmetric field theories \cite{BK0,SGRS}.

Here in this letter, we would like to consider a nonlocal extension of the chiral superfield model in a four-dimensional spacetime.  Within this study, we restrict ourselves to Lorentz-invariant theories only. So, we start with the general model below: 
\bea
\label{sfigeral}
S_{HD}=\int d^8z\bar\Phi h(D^2,\bar{D}^2,\Box){\Phi}+\Big\{\int d^6z\bigg[\frac{m}{2}\Phi g(\Box)\Phi+\frac{\lambda}{3!}\Phi^3\bigg]+h.c.\Big\} \ ,
\eea
where the derivatives are introduced in the Wess-Zumino model through the dimensionless scalar operators $h(D^2,\bar{D}^2,\Box)$ and $g(\Box)$. We assume that the functions $h$ and $g$ are analytical functions of the squared supercovariant derivatives. Note that, we choose $g(\Box)$ to be a function only of the D'Alembertian operator to enforce the integrand in the second term to be chiral. 
We note that the model in (\ref{sfigeral}) can be simplified by expanding $h(D^2,\bar{D}^2,\Box)$ in Taylor series with respect to the spinor covariant derivatives and using the well-known properties of the chiral superfields \cite{BK0,SGRS}. Thus, it can be shown that
\bea
\label{ident}
\bar{\Phi}h(D^2,\bar{D}^2,\Box)\Phi=\bar{\Phi}f(\Box)\Phi+D^2(\bar{\Phi}\ell(\Box)\Phi) \ .
\eea
Substituting (\ref{ident}) into (\ref{sfigeral}), we obtain
\bea
\label{sfigeral2}
S_{HD}=\int d^8z\bar\Phi f(\Box){\Phi}+\Big\{\int d^6z\bigg[\frac{m}{2}\Phi g(\Box)\Phi+\frac{\lambda}{3!}\Phi^3\bigg]+h.c.\Big\} \ ,
\eea
where we have neglected the surface term. We notice that the model (\ref{sfigeral}) is equivalent to the one (\ref{sfigeral2}). Thus, from now on, all the calculations presented in this work will be performed for (\ref{sfigeral2}). Four different non-local models will be considered. Typically, except  of the Model I which admits some essentially distinct motivation, we will suggest that $f(\Box)$ and $g(\Box)$ are analytic functions so that the limit of $f(z)$ and $g(z)$ as $z\to0$ coincide with the identity.

The component form of the above action is given by
\bea
\label{componentes}
S_{HD}&=&\int d^4x\left[\bar{A}\Box f(\Box)A+\bar\psi^{\dot{\alpha}}f(\Box)i{\partial^\alpha}_{\dot{\alpha}}\psi_\alpha+\bar{F}f(\Box)F\right]\nonumber\\
&+&\int d^4x\bigg\{\frac{m}{2}\big[Fg(\Box)A+\psi^\alpha g(\Box)\psi_\alpha+A g(\Box)F\big]+\lambda(A\psi^2+\frac{1}{2}FA^2)+h.c.\bigg\} \ .
\eea 

Regarding the quantum aspects of the model (\ref{sfigeral2}), the Feynman rules for such model are similar to those ones the usual chiral superfield model, except that the new propagators are given by
\bea
\label{prop1}
&&\langle\Phi(-k,\theta_1)\bar{\Phi}(k,\theta_2)\rangle=\frac{f(-k^2)}{f^2(-k^2)k^2+g^2(-k^2) m^2}\delta^4(\theta_1-\theta_2) \ ; \\
\label{prop2}
&&\langle\Phi(-k,\theta_1)\Phi(k,\theta_2)\rangle=\langle\bar{\Phi}(-k,\theta_1)\bar{\Phi}(k,\theta_2)\rangle^*=-
\frac{g(-k^2)m D^2}{k^2\big[f^2(-k^2)k^2+g^2(-k^2)m^2\big]}\delta^4(\theta_1-\theta_2) \  .
\eea
This difference occurs due to the fact that the higher-derivative modification of the chiral superfield action occurs only in the quadratic terms in (\ref{sfigeral2}).

Equations (\ref{prop1}) and (\ref{prop2}) make explicit that adding higher-derivative terms to the chiral superfield action, through a suitable choice of the operators $f(\Box)$ and $g(\Box)$, improves the
ultraviolet behavior of the theory. Indeed, it is old and well-known fact that higher-derivative theories have better ultraviolet behavior than conventional ones \cite{LWS}. On the other hand, higher-derivative theories are accompanied by some degree of skepticism about their physical viability. In order to illustrate the problem with these theories, let us assume that $f(\Box)=g(\Box)$ in (\ref{sfigeral2}). Thus, Eqs. (\ref{prop1}) and (\ref{prop2}) can be rewritten as
\bea
\label{prop3}
&&\langle\Phi(-k,\theta_1)\bar{\Phi}(k,\theta_2)\rangle=\frac{1}{f(-k^2)\big(k^2+m^2\big)}\delta^4(\theta_1-\theta_2) \ ; \\
\label{prop4}
&&\langle\Phi(-k,\theta_1)\Phi(k,\theta_2)\rangle=-
\frac{m D^2}{f(-k^2)k^2\big(k^2+m^2\big)}\delta^4(\theta_1-\theta_2) \  .
\eea
Now, consider a higher-derivative version of the chiral superfield model, where the the higher-derivative operator is given by $f(\Box)=-\xi\Box+1$. Thus, it follows from (\ref{prop3}-\ref{prop4}) that the propagators for this model are given by
\bea
\label{prop5}
&&\langle\Phi(-k,\theta_1)\bar{\Phi}(k,\theta_2)\rangle=\frac{1}{1-\xi m^2}\bigg[\frac{1}{k^2+m^2}-\frac{1}{k^2+\xi^{-1}}\bigg]\delta^4(\theta_1-\theta_2) \ ; \\
\label{prop6}
&&\langle\Phi(-k,\theta_1)\Phi(k,\theta_2)\rangle=-\frac{m}{1-\xi m^2}\bigg[\frac{1}{k^2+m^2}-\frac{1}{k^2+\xi^{-1}}\bigg]
\frac{D^2}{k^2}\delta^4(\theta_1-\theta_2) \  .
\eea
Note that the higher-derivative term introduces a new degree of freedom represented by a new pole into the propagator at $k^2=-\xi^{-1}$, but due to the fact that such massive propagator has the "wrong" sign, the new degree of freedom will contribute to the Hamiltonian with negative kinetic energy. Thus, the Hamiltonian for this theory is not bounded from below, so that the theory becomes
unstable.  Unfortunately, this problem plagues all higher-derivative theories described by the Ostrogradsky's theorem \cite{Ghost}.

Alternatively, the above problem can be avoided if we consider a version of the chiral superfield model with infinitely many derivatives \cite{NLSUSY}. Indeed, Eqs. (\ref{prop3}-\ref{prop4}) suggest that, if $f(\Box)$ is an entire function with no zeros, then the pole structure of the propagator is the same as that of the usual chiral superfield model. Additionally, if $f(\Box)$ is an entire function with no zeros, then there is an entire function $h(\Box)$ such that $f(\Box)=e^{h(\Box)}$ \cite{Vaughn}. Therefore, through a suitable choice of the operator $h(\Box)$, we can construct a UV-finite theory with infinitely many derivatives  without introducing unphysical degrees of freedom \cite{Efimov,BMS}.

This paper is organized as follows. In Sec. \ref{1loop} we introduce 
the nonlocal superfield action and provide a complete general expression for the one-loop effective potential. In Sec. \ref{NL} we calculate the one-loop correction to the potential considering four different nonlocal models. In this way we demonstrate how the non-locality, whose intensity is characterized by the parameter $\Lambda$ describing the characteristic energy at which the nonlocality becomes significant, modifies the effective potential when compared with the standard Wess-Zumino one. Finally we leave to Sec. \ref{Conc} our conclusions and most relevant remarks.

\section{One-loop effective potential}
\label{1loop}

Let us now calculate the one-loop  contribution, $K^{(1)}(\bar\Phi,\Phi)$, for the  K\"{a}hlerian effective potential of the higher-derivative model (\ref{sfigeral2}).  Following the general approach (cf. \cite{BK0}), firstly, we split the superfields $\Phi\to\Phi+\phi$ and $\bar\Phi\to\bar\Phi+\bar\phi$, where $\Phi,\bar{\Phi}$ are background superfields, and $\phi,\bar{\phi}$ are quantum ones. Secondly, we expand (\ref{sfigeral2}) around the background superfields. Therefore, it follows from (\ref{sfigeral2}) that
\bea
\label{quadratic}
S_2[\bar\Phi,\Phi;\bar\phi,\phi]=\int d^8z\bar\phi f(\Box){\phi}+\frac{1}{2}\Big\{\int d^6z\phi\big(g(\Box)m+\lambda\Phi\big)\phi+h.c.\Big\} \  ,
\eea 
where we have kept only the quadratic terms in the quantum superfields for the one-loop calculations.

For convenience, we will extract the propagators from the general Lagrangian and treat the antichiral and chiral Lagrangians as interaction vertices in (\ref{quadratic}). Thus, it follows that the propagator is given by
\bea
\label{quantumprop}
\langle\phi(-k,\theta_1)\bar{\phi}(k,\theta_2)\rangle=\frac{1}{k^2f(-k^2)}\delta^4(\theta_1-\theta_2) \ .
\eea
 The $D$-factors within our study are associated with vertices just by the same rules as in the usual Wess-Zumino model \cite{BK0}.

Equipped with the propagator and vertices (\ref{quadratic}-\ref{quantumprop}), we can infer that the sequence of supergraphs below are non-vanishing. In these supergraphs, the internal lines are the propagators $<\phi(-k)\bar{\phi}(k)>$, and the external ones are for $g(-k^2)m+\lambda\Phi$.

\hspace{3.5cm}
\unitlength=.6mm
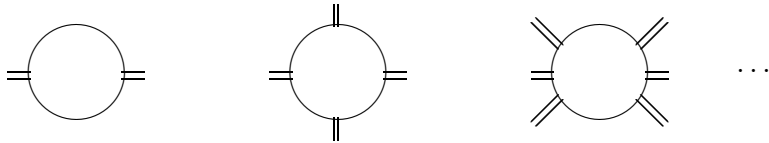
\begin{picture}(20,20)
\put(0,10){\circle{20}}\put(-10,10){\line(-1,0){5}}\put(-10,8.5){\line(-1,0){5}}
\put(10,10){\line(1,0){5}}\put(10,8.5){\line(1,0){5}}
%\ind(24,0){W''}\ind(-26,0){\bar{W''}}
%\put(4.8,0){\GRAPH(hsize=3){
\end{picture}
\hspace{2cm}
\begin{picture}(20,20)
\put(0,10){\circle{20}}\put(-10,10){\line(-1,0){5}}
\put(-10,8.5){\line(-1,0){5}}
\put(10,10){\line(1,0){5}}\put(10,8.5){\line(1,0){5}}
\put(0,20){\line(0,1){5}}\put(-1,20){\line(0,1){5}}
\put(0,0){\line(0,-1){5}}
\put(-1,0){\line(0,-1){5}}
%\put(0,-13){Fig.1}
%}}
%\ind(-10,20){W''}\ind(-10,-20){W''}
%\ind(-49,0){\bar{W}''}\ind(5,0){\bar{W''}}
%\put(30,0){\ldots}
\end{picture}
%}}}}
\hspace{2cm}
\begin{picture}(30,30)
\put(0,10){\circle{20}}
\put(-10,10){\line(-1,0){5}}\put(-10,8.5){\line(-1,0){5}}
\put(10,10){\line(1,0){5}}\put(10,8.5){\line(1,0){5}}
\put(-9,5){\line(-1,-1){6}}\put(-8,4){\line(-1,-1){6}}
\put(9,5){\line(1,-1){6}}\put(8,4){\line(1,-1){6}}
\put(-9,15){\line(-1,1){6}}\put(-8,16){\line(-1,1){6}}
\put(9,15){\line(1,1){6}}\put(8,16){\line(1,1){6}}
\put(30,10){\ldots}
\end{picture}
\hspace{3.5cm}

The complete sum of these supergraphs can be calculated as usual {(cf. \cite{Grisaru96})}. For the model (\ref{sfigeral2}), it yields
\bea
\label{genericeffecPot}
K^{(1)}(\bar\Phi,\Phi)=\frac{1}{2(4\pi)^2}\int_0^\infty dk^2\ln\left\{1+\frac{\big[g(-k^2)m+\lambda\Phi\big]\big[g(-k^2)m+\lambda\bar\Phi\big]}{k^2f^2(-k^2)}\right\} \ .
\eea 
In order to solve the integral above, we must choose the explicit form of the functions $f(\Box)$ and $g(\Box)$ in (\ref{sfigeral2}). Unfortunately, even for the simplest choices of the nonlocal operators $f(\Box)$ and $g(\Box)$, it is not possible to integrate  over $k^2$ exactly and obtain a closed expression  for \eqref{genericeffecPot}. Thus, let us consider four nonlocal models and provide an approximated expression for the effective potential of each model.

\section{Non-local models}
\label{NL}
Here in this section we calculate the one-loop effective potential for this theory.
To do that we consider the general expression given in \eqref{genericeffecPot}, considering explicitly four different non-local models. So, in what follows we make this calculation explicitly in the case of some specific choices of the function $f(z)$ and $g(z)$. All but the first model reduce to the Wess-Zumino model when we take the non-local parameter  to be infinitely large, $\Lambda\to\infty$.

\subsection{Model I}

Our first model is described by the operators $f(\Box)=-\xi^2\Box^{-1}e^{-\frac{\Box}{\Lambda^2}}$ and $g(\Box)=0$, where $\xi$ and $\Lambda$ are energy scale parameters. If we substitute these functions into Eq. (\ref{componentes}), we would notice that this model has a kinetic term similar to that of the effective Lagrangian for the p-adic string derived in \cite{BFOW}. On the other hand, we choose this model because, among the models studied here, it produces the simplest expression of the integral (\ref{genericeffecPot}), which is
\bea
\label{model1}
K^{(1)}_I(\bar\Phi,\Phi)=\frac{\Lambda^2}{(8\pi)^2}\int_0^\infty dk^2\ln\left\{1+\frac{2\Lambda^2\left|\lambda\Phi\right|^2}{\xi^4}k^2e^{-k^2}\right\} \ ,
\eea 
where we rescale the integration variable by $k^2\rightarrow\frac{\Lambda^2}{2}k^2$. To solve this integral, we make the assumption that $\Lambda^2\left|\lambda\Phi\right|^2\ll\xi^4$, so that we can expand the logarithm in a power series and obtain
\bea
\label{series1}
K^{(1)}_I(\bar\Phi,\Phi)&=&-\frac{\Lambda^2}{(8\pi)^2}\sum_{n=1}^\infty\left(-\frac{2\Lambda^2\left|\lambda\Phi\right|^2}{\xi^4}\right)^n \int_0^\infty dk^2\left(k^2\right)^n e^{-nk^2}\nonumber\\
&=&-\frac{\Lambda^2}{(8\pi)^2}\sum_{n=1}^\infty\left(-\frac{2\Lambda^2\left|\lambda\Phi\right|^2}{n\xi^4}\right)^n \Gamma(n) \ ,
\eea 
or
\bea
\label{series_1}
K^{(1)}_I(\bar\Phi,\Phi)\approx\frac{2\Lambda^4\left|\lambda\Phi\right|^2}{(8\pi)^2\xi^4}\left[1-\frac{\Lambda^2\left|\lambda\Phi\right|^2}{2\xi^4}+\frac{8\Lambda^4\left|\lambda\Phi\right|^4}{27\xi^{8}}-{\cal O}\left(\frac{\Lambda^6\left|\lambda\Phi\right|^6}{\xi^{12}}\right)\right] \ ,
\eea 
where we write the first three terms of the series explicitly. So, this model presents a strong non-locality.

\subsection{ Model II}

For the second model, we choose the operators $f(\Box)=e^{-\frac{\Box}{\Lambda^2}}$ and $g(\Box)=1$, where $\Lambda$ is an energy scale. In particular, when we take $\Lambda\rightarrow\infty$, we recover the Wess-Zumino model. These operators were considered in \cite{Briscese15}, where the one-loop effective potential for a nonlocal scalar field theory was calculated. Thus, it follows from (\ref{genericeffecPot}) that
\bea
\label{model2}
K^{(1)}_{II}(\bar\Phi,\Phi)=\frac{1}{2(4\pi)^2}\int_0^\infty dk^2\ln\left[1+\frac{\left|\Psi\right|^2}{k^2}e^{-2\frac{k^2}{\Lambda^2}}\right] \ ,
\eea
where $\left|\Psi\right|^2\equiv\left(m+\lambda\Phi\right)\left(m+\lambda\bar\Phi\right)$.

In order to evaluate this integral, we assume that $\left|\Psi\right|^2\ll\Lambda^2$ (we note that in the paper \cite{Trodden}, the scale $\Lambda$ at which the higher derivatives, and, as a consequence, the  non-locality, become essential, is supposed to be of the order of the Planck mass). However, we cannot proceed as in
the previous model, because an expansion in powers of $\frac{\left|\Psi\right|^2}{\Lambda^2}$ gives rise to IR divergent integrals. Hence, we will solve (\ref{model2}) by applying the  strategy of expansion by regions \cite{BS,Smirnov}.

Let us divide the interval of integration in two subintervals; that is $[0,\Omega^2]$ and $[\Omega^2,\infty)$, where we introduce an intermediate scale $\Omega^2$ satisfying $\left|\Psi\right|^2\ll\Omega^2\ll\Lambda^2$ \cite{BBF}. Thereby, we can split the integral (\ref{model2}) into two parts
\bea
\label{split2}
K^{(1)}_{II}(\bar\Phi,\Phi)&=&\frac{1}{32\pi^2}\left\{\int_0^{\Omega^2} dk^2\ln\left[1+\frac{\left|\Psi\right|^2}{k^2}e^{-2\frac{k^2}{\Lambda^2}}\right]+\int_{\Omega^2}^\infty dk^2\ln\left[1+\frac{\left|\Psi\right|^2}{k^2}e^{-2\frac{k^2}{\Lambda^2}}\right]\right\}\nonumber\\
&\equiv&\frac{1}{32\pi^2}\left\{I_L+I_H\right\} \ .
\eea 
On the one hand, in the low-energy region $[0,\Omega^2]$, we note that $k^2\sim\left|\Psi\right|^2\ll\Lambda^2$. Therefore, we can expand the first integrand on the r.h.s. of (\ref{split2}) and obtain
\bea
\label{I_L-2}
I_L=\int_0^{\Omega^2} dk^2\Bigg\{\ln\left[1+\frac{\left|\Psi\right|^2}{k^2}\right]-2\frac{\left|\Psi\right|^2}{\Lambda^2}\frac{k^2}{k^2+\left|\Psi\right|^2}+2\frac{\left|\Psi\right|^2}{\Lambda^4}\frac{(k^2)^3}{\big(k^2+\left|\Psi\right|^2\big)^2}+\cdots\Bigg\} \ ,
\eea
where we have retained terms up to the second order in $1/\Lambda^2$.

On the other hand, in the high-energy region $[\Omega^2,\infty)$, we note that $\left|\Psi\right|^2\ll k^2\sim\Lambda^2$. Therefore, we can expand the second integrand on the r.h.s. of (\ref{split2}) and obtain
\bea
\label{I_H-2}
I_H=\int_{\Omega^2}^\infty dk^2\Bigg\{\left|\Psi\right|^2\frac{e^{-2\frac{k^2}{\Lambda^2}}}{k^2}-\frac{\left|\Psi\right|^4}{2}\frac{e^{-4\frac{k^2}{\Lambda^2}}}{k^4}+\frac{\left|\Psi\right|^6}{3}\frac{e^{-6\frac{k^2}{\Lambda^2}}}{k^6}+\cdots\Bigg\} \ ,
\eea
where we have kept terms up to the third order in $\left|\Psi\right|^2$.

The above integrals can be solved in a closed form. However, to study the asymptotic behavior it is sufficient to calculate only their approximate expressions. Hence, we can integrate Eq. (\ref{I_L-2}) to yield
\bea
\label{aproxI_L-2}
I_L&\approx&\left|\Psi\right|^2\left\{1-\ln\left(\frac{\left|\Psi\right|^2}{\Omega^2}\right)+\frac{\left|\Psi\right|^2}{2\Omega^2}-\frac{\left|\Psi\right|^4}{6\Omega^4}\right\}-\frac{2\left|\Psi\right|^2}{\Lambda^2}\left\{\Omega^2+\left|\Psi\right|^2\left[\ln\left(\frac{\left|\Psi\right|^2}{\Omega^2}\right)-\frac{\left|\Psi\right|^2}{\Omega^2}\right]\right\}\nonumber\\
&+&\frac{2\left|\Psi\right|^2}{\Lambda^4}\left\{\frac{\Omega^4}{2}-2\left|\Psi\right|^2\Omega^2-\left|\Psi\right|^4\left[1+3\ln\left(\frac{\left|\Psi\right|^2}{\Omega^2}\right)\right]\right\} \ .
\eea
Similarly, we get from Eq. (\ref{I_H-2}) the result
\bea
\label{aproxI_H-2}
I_H&\approx&\left|\Psi\right|^2\left\{-\gamma-\ln\left(\frac{2\Omega^2}{\Lambda^2}\right)+\frac{2\Omega^2}{\Lambda^2}-\frac{\Omega^4}{\Lambda^4}\right\}-\frac{\left|\Psi\right|^4}{2}\left\{\frac{1}{\Omega^2}+\frac{4}{\Lambda^2}\left[-1+\gamma+\ln\left(\frac{4\Omega^2}{\Lambda^2}\right)-\frac{2\Omega^2}{\Lambda^2}\right]\right\}\nonumber\\
&+&\frac{\left|\Psi\right|^6}{3}\left\{\frac{1}{2\Omega^4}-\frac{6}{\Lambda^2\Omega^2}-\frac{9}{\Lambda^4}\left[-3+2\gamma+2\ln\left(\frac{6\Omega^2}{\Lambda^2}\right)\right]\right\} \ .
\eea
Finally, substituting (\ref{aproxI_L-2}) and (\ref{aproxI_H-2}) into (\ref{split2}), we get
\bea
\label{series2}
K^{(1)}_{II}(\bar\Phi,\Phi)\approx-\frac{\left|\Psi\right|^2}{32\pi^2}\left\{\ln\left(\frac{2\left|\Psi\right|^2}{e^{1-\gamma}\Lambda^2}\right)+\frac{2\left|\Psi\right|^2}{\Lambda^2}\ln\left(\frac{4\left|\Psi\right|^2}{e^{1-\gamma}\Lambda^2}\right)+\frac{\left|\Psi\right|^4}{\Lambda^4}\left[-1+6\ln\left(\frac{6\left|\Psi\right|^2}{e^{1-\gamma}\Lambda^2}\right)\right]\right\},
\eea
where $\left|\Psi\right|^2\equiv\left(m+\lambda\Phi\right)\left(m+\lambda\bar\Phi\right)$.

Note that the artificial scale $\Omega^2$ is completely cancelled in the sum of (\ref{aproxI_L-2}) and (\ref{aproxI_H-2}) to produce the result (\ref{series2}). Indeed, this had to occur, because $\Omega^2$  is not present in the integral (\ref{model2}). Moreover, we note that, taking the limit $\Lambda\rightarrow\infty$ in (\ref{series2}) with a replacement $e^{1-\gamma}\Lambda^2\to\mu^2$, the one-loop effective potential for the Wess-Zumino model is reproduced \cite{BKY}. In this particular case, the scale $\Lambda$ plays the role of an ultraviolet cutoff scale.

\subsection{ Model III}

Our third model is characterized by the operators $f(\Box)=g(\Box)=e^{-\frac{\Box}{\Lambda^2}}$.  As the previous model, this model also reduces to the Wess-Zumino model in the limit where the scale $\Lambda$ becomes infinite. However, unlike the previous model, the propagators of this model have the same pole structure as the ones of the Wess-Zumino model [see Eqs. (\ref{prop1}) and (\ref{prop2})].

Substituting $f(\Box)=g(\Box)=e^{-\frac{\Box}{\Lambda^2}}$ into (\ref{genericeffecPot}), we get
\bea
\label{model3}
K^{(1)}_{III}(\bar\Phi,\Phi)=\frac{1}{2(4\pi)^2}\int_0^\infty dk^2\ln\left[1+\frac{1}{k^2}\left(\alpha+\beta e^{-\frac{k^2}{\Lambda^2}}+\delta e^{-2\frac{k^2}{\Lambda^2}}\right)\right] \ ,
\eea
where $\alpha\equiv m^2$, $\beta\equiv m\lambda(\bar\Phi+\Phi)$, and $\delta\equiv\lambda^2\bar\Phi\Phi$.

Again, let us apply the strategy of regions to solve this integral. We assume that $\alpha,\beta,\delta\ll\Lambda^2$, so that we can introduce $\Omega^2$ satisfying $\alpha,\beta,\delta\ll\Omega^2\ll\Lambda^2$. Therefore, we can rewrite (\ref{model3}) as
\bea
\label{split3}
K^{(1)}_{III}(\bar\Phi,\Phi)&=&\frac{1}{32\pi^2}\bigg\{\int_0^{\Omega^2} dk^2\ln\left[1+\frac{1}{k^2}\left(\alpha+\beta e^{-\frac{k^2}{\Lambda^2}}+\delta e^{-2\frac{k^2}{\Lambda^2}}\right)\right]+\int_{\Omega^2}^\infty dk^2\ln\bigg[1+\frac{1}{k^2}\bigg(\alpha\nonumber\\
&+&\beta e^{-\frac{k^2}{\Lambda^2}}+\delta e^{-2\frac{k^2}{\Lambda^2}}\bigg)\bigg]\bigg\}\equiv\frac{1}{32\pi^2}\left\{I_L+I_H\right\} \ .
\eea 
On the one hand, $k^2\sim\alpha,\beta,\delta\ll\Lambda^2$  in the low-energy region. Therefore, we will expand the first integrand on the r.h.s. of (\ref{split3}) in powers of $1/\Lambda^2$ and keep only the first two terms:
\bea
\label{I_L-3}
I_L=\int_0^{\Omega^2} dk^2\left\{\ln\left[1+\frac{\alpha+\beta+\delta}{k^2}\right]-\frac{\left(\beta+2\delta\right)}{\Lambda^2}\frac{k^2}{k^2+\alpha+\beta+\delta}+\cdots\right\} \ .
\eea
On the other hand, $\alpha,\beta,\delta\ll k^2\sim\Lambda^2$ in the high-energy region. Therefore, we can expand the second integrand on the r.h.s. of (\ref{split3}) in powers of $\alpha,\beta,\delta$ and retain only the terms up to second order in $\alpha,\beta,\delta$:
\bea
\label{I_H-3}
I_H=\int_{\Omega^2}^\infty dk^2\bigg\{\delta\frac{e^{-2\frac{k^2}{\Lambda^2}}}{k^2}-\left(\alpha\delta+\frac{\beta^2}{2}\right)\frac{e^{-2\frac{k^4}{\Lambda^2}}}{k^4}-\beta\delta\frac{e^{-3\frac{k^2}{\Lambda^2}}}{k^4}-\delta^2\frac{e^{-4\frac{k^2}{\Lambda^2}}}{2k^4}+\cdots\bigg\} \ .
\eea
Notice that we have discarded terms proportional to $\alpha$, $\alpha^2$, $\beta$, and $\alpha\beta$ in (\ref{I_H-3}). Indeed, since these terms can be expressed in the form $F(\Phi)+\bar{F}(\bar{\Phi})$, and these terms identically vanish being integrated over the whole superspace, it follows that $\alpha$, $\alpha^2$, $\beta$, and $\alpha\beta$ give trivial contributions to the effective action, so that they can be omitted. Additionally, we will take advantage of this invariance of the effective action, and at the end of our calculation, we will conveniently add to (\ref{split3}) the trivial expression
\bea
\label{aproxI_T-3}
I_T\equiv-\frac{1}{32\pi^2}\left\{(\alpha+\beta)\ln\left(\frac{2\Omega^2}{e^{-\gamma}\Lambda^2}\right)+\frac{\alpha\beta}{\Lambda^2}\ln\left(\frac{\Omega^2}{e^{1-\gamma}\Lambda^2}\right)\right\} \ .
\eea
We notice that Eqs. (\ref{I_L-3}) and (\ref{I_H-3}) can be integrated to yield
\bea
\label{aproxI_L-3}
I_L&\approx&(\alpha+\beta+\delta)\left\{1-\ln\left(\frac{\alpha+\beta+\delta}{\Omega^2}\right)+\frac{\alpha+\beta+\delta}{2\Omega^2}\right\}-\frac{(\beta+2\delta)}{\Lambda^2}\bigg\{\Omega^2+(\alpha+\beta+\delta)\nonumber\\
&\times&\ln\left(\frac{\alpha+\beta+\delta}{\Omega^2}\right)\bigg\}\ ,
\eea
and
\bea
\label{aproxI_H-3}
I_H&\approx&\delta\left\{-\gamma-\ln\left(\frac{2\Omega^2}{\Lambda^2}\right)+\frac{2\Omega^2}{\Lambda^2}\right\}-\left(\alpha\delta+\frac{\beta^2}{2}\right)\left\{\frac{1}{\Omega^2}+\frac{2}{\Lambda^2}\left[\gamma-1+\ln\left(\frac{2\Omega^2}{\Lambda^2}\right)\right]\right\}-\beta\delta\bigg\{\frac{1}{\Omega^2}\nonumber\\
&+&\frac{3}{\Lambda^2}\left[\gamma-1+\ln\left(\frac{3\Omega^2}{\Lambda^2}\right)\right]\bigg\}-\frac{\delta^2}{2}\left\{\frac{1}{\Omega^2}+\frac{4}{\Lambda^2}\left[\gamma-1+\ln\left(\frac{4\Omega^2}{\Lambda^2}\right)\right]\right\} \ .
\eea
Finally, by substituting Eqs. (\ref{aproxI_L-3}) and (\ref{aproxI_H-3}) into (\ref{split3}), and adding (\ref{aproxI_T-3}) to the final result, we obtain
\bea
\label{series3}
K^{(1)}_{III}(\bar\Phi,\Phi)&\approx&-\frac{1}{32\pi^2}\bigg\{(\alpha+\beta+\delta)\ln\left(\frac{2(\alpha+\beta+\delta)}{e^{1-\gamma}\Lambda^2}\right)+\frac{1}{\Lambda^2}\bigg[\alpha\beta\ln\left(\frac{\alpha+\beta+\delta}{e^{1-\gamma}\Lambda^2}\right)+(\beta^2+2\alpha\delta)\nonumber\\
&\times&\ln\left(\frac{2(\alpha+\beta+\delta)}{e^{1-\gamma}\Lambda^2}\right)+3\beta\delta\ln\left(\frac{3(\alpha+\beta+\delta)}{e^{1-\gamma}\Lambda^2}\right)+2\delta^2\ln\left(\frac{4(\alpha+\beta+\delta)}{e^{1-\gamma}\Lambda^2}\right)\bigg]\bigg\} \ ,
\eea
where $\alpha\equiv m^2$, $\beta\equiv m\lambda(\bar\Phi+\Phi)$, and $\delta\equiv\lambda^2\bar\Phi\Phi$.

Just as the effective potential (\ref{series2}), the result (\ref{series3}) also does not depend on the scale $\Omega^2$ and reproduces the renormalized one-loop effective potential for the Wess-Zumino model in the limit $\Lambda\rightarrow\infty$,  where again the nonlocality parameter $\Lambda$ plays the role of the normalization parameter $\mu$.  We see that the  leading (logarithmic) orders of the expressions (\ref{series2},\ref{series3}) identically coincide.

\subsection{ Model IV}

Our last model is described by the operators $f(\Box)=1$ and $g(\Box)=e^{\frac{\Box}{\Lambda^2}}$. This model is characterized by the presence of higher derivatives only in the chiral sector, as can be seen in Eq. (\ref{sfigeral2}), so that it is similar to the local higher-derivative chiral superfield model studied in Ref. \cite{GGNPS}, but it is different from the ones studied so far in this letter.

We can substitute $f(\Box)=1$ and $g(\Box)=e^{\frac{\Box}{\Lambda^2}}$ into (\ref{genericeffecPot}) to obtain
\bea
\label{model4}
K^{(1)}_{IV}(\bar\Phi,\Phi)=\frac{1}{2(4\pi)^2}\int_0^\infty dk^2\ln\left[1+\frac{1}{k^2}\left(\delta+\beta e^{-\frac{k^2}{\Lambda^2}}+\alpha e^{-2\frac{k^2}{\Lambda^2}}\right)\right] \ ,
\eea
where $\alpha$, $\beta$, and $\delta$ are as defined in (\ref{model3}). It should be observed that integrals (\ref{model3}) and (\ref{model4}) are quite similar (indeed, the Eq. (\ref{model4}) can be obtained from the Eq. (\ref{model3}) through the replacement $\alpha\leftrightarrow\delta$) . However, as we will see, the latter does not lead to a finite effective potential. For this reason, let us use dimensional regularization to regulate the divergent integral above, so that we can replace the four-dimensional integral (\ref{model4}) by the $d$-dimensional one:
\bea
\label{regmodel4}
K^{(1)}_{IV}(\bar\Phi,\Phi)=\frac{\mu^{2\varepsilon}}{2(4\pi)^{2-\varepsilon}\Gamma(2-\varepsilon)}\int_0^\infty dk^2(k^2)^{-\varepsilon}\ln\left[1+\frac{1}{k^2}\left(\delta+\beta e^{-\frac{k^2}{\Lambda^2}}+\alpha e^{-2\frac{k^2}{\Lambda^2}}\right)\right] \ ,
\eea
where $\mu$ is an arbitrary mass scale and $\varepsilon=2-\frac{d}{2}\rightarrow 0$.

Let us assume that $\alpha,\beta,\delta\ll\Lambda^2$, so that we can evaluate (\ref{regmodel4}) by applying the strategy of regions. Therefore, it follows from (\ref{regmodel4}) that
\bea
\label{split4}
K^{(1)}_{IV}(\bar\Phi,\Phi)=\frac{1}{32\pi^2}\left\{I_L+I_H\right\}+I_T \ ,
\eea 
where
\bea
\label{I_L-4}
I_L&=&\frac{(4\pi\mu^2)^{\varepsilon}}{\Gamma(2-\varepsilon)}\int_0^{\Omega^2} dk^2(k^2)^{-\varepsilon}\left\{\ln\left[1+\frac{\alpha+\beta+\delta}{k^2}\right]-\frac{\left(\beta+2\alpha\right)}{\Lambda^2}\frac{k^2}{k^2+\alpha+\beta+\delta}+\cdots\right\} \ ; \\
\label{I_H-4}
I_H&=&\frac{(4\pi\mu^2)^{\varepsilon}}{\Gamma(2-\varepsilon)}\int_{\Omega^2}^\infty dk^2(k^2)^{-\varepsilon}\bigg\{\frac{\delta}{k^2}-\frac{\delta^2}{2k^4}-\left(\alpha\delta+\frac{\beta^2}{2}\right)\frac{e^{-2\frac{k^4}{\Lambda^2}}}{k^4}-\beta\delta\frac{e^{-\frac{k^2}{\Lambda^2}}}{k^4}+\cdots\bigg\} \ ; \\
\label{I_T-4}
I_T&\equiv&-\frac{1}{32\pi^2\Lambda^2}\left[2\alpha^2\ln\left(\frac{16\pi\mu^2}{\Lambda^2}\right)+3\alpha\beta\ln\left(\frac{12\pi\mu^2}{\Lambda^2}\right)\right] \ .
\eea
As a matter of convenience, we have added to (\ref{split4}) the trivial contribution $I_T$ and discarded terms proportional to $\alpha$, $\alpha^2$, $\beta$, and $\alpha\beta$ in (\ref{I_H-4}).

The above integrals are more easily evaluated if we integrate over the full integration domain $k^2\in[0,\infty)$.  Thus, in order to simplify our calculations, let us split the integrals (\ref{I_L-4}-\ref{I_H-4}) into
\bea
\label{SI_L-4}
I_L&=&\frac{(4\pi\mu^2)^{\varepsilon}}{\Gamma(2-\varepsilon)}\left[\int_{0}^\infty dk^2-\int_{\Omega^2}^\infty dk^2\right](k^2)^{-\varepsilon}\bigg\{\ln\left[1+\frac{\alpha+\beta+\delta}{k^2}\right]-\frac{\left(\beta+2\alpha\right)}{\Lambda^2}\nonumber\\
&\times&\frac{k^2}{k^2+\alpha+\beta+\delta}+\cdots\bigg\}\equiv I_L\big|_{\Omega^2\rightarrow\infty}-R_L \ ,
\eea
and
\bea
\label{SI_H-4}
I_H&=&\frac{(4\pi\mu^2)^{\varepsilon}}{\Gamma(2-\varepsilon)}\left[\int_{0}^\infty dk^2-\int_{0}^{\Omega^2} dk^2\right](k^2)^{-\varepsilon}\bigg\{\frac{\delta}{k^2}-\frac{\delta^2}{2k^4}-\left(\alpha\delta+\frac{\beta^2}{2}\right)\frac{e^{-2\frac{k^4}{\Lambda^2}}}{k^4}\nonumber\\
&-&\beta\delta\frac{e^{-\frac{k^2}{\Lambda^2}}}{k^4}+\cdots\bigg\}\equiv I_H\big|_{\Omega^2\rightarrow 0}-R_H \ .
\eea
The integrals on the interval $k^2\in[0,\infty)$ are well-known and can be computed easily. Therefore, we obtain
\bea
\label{LI_L-4}
I_L\big|_{\Omega^2\rightarrow\infty}&=&\left(\alpha+\beta+\delta\right)\left[\frac{1}{\varepsilon}+2-\gamma-\ln\left(\frac{\alpha+\beta+\delta}{4\pi\mu^2}\right)\right]+\frac{\left(\beta+2\alpha\right)\left(\alpha+\beta+\delta\right)}{\Lambda^2}\bigg[\frac{1}{\varepsilon}+1\nonumber\\
&-&\gamma-\ln\left(\frac{\alpha+\beta+\delta}{4\pi\mu^2}\right)\bigg] \ ;\\
\label{LI_H-4}
I_H\big|_{\Omega^2\rightarrow 0}&=&-\frac{1}{\Lambda^2}\delta\beta\left[\frac{1}{\varepsilon}+\ln\left(\frac{4\pi\mu^2}{\Lambda^2}\right)\right]-\frac{1}{\Lambda^2}\left(\beta^2+2\alpha\delta\right)\left[\frac{1}{\varepsilon}+\ln\left(\frac{8\pi\mu^2}{\Lambda^2}\right)\right] \ .
\eea
On the other hand, the integrals in $R_L$ and $R_H$ do not need to be explicitly calculated because $R_L+R_H=0$. In order to prove this statement, we notice that $R_L$ can be expanded in powers of $\alpha,\beta,\delta$ and retained only the terms up to second order in $\alpha,\beta,\delta$:
\bea
\label{R_L-4}
R_L&=&\frac{(4\pi\mu^2)^{\varepsilon}}{\Gamma(2-\varepsilon)}\int_{\Omega^2}^\infty dk^2(k^2)^{-\varepsilon}\bigg\{\left(-\frac{1}{2k^4}+\frac{1}{\Lambda^2 k^2}\right)\beta^2+\left(-\frac{1}{k^4}+\frac{2}{\Lambda^2 k^2}\right)\alpha\delta+\bigg(-\frac{1}{k^4}\nonumber\\
&+&\frac{1}{\Lambda^2 k^2}\bigg)\beta\delta+\frac{\delta}{k^2}-\frac{\delta^2}{2k^4}+\cdots\bigg\} \ .
\eea
At the same time, we can also expand $R_H$ in powers of $1/\Lambda^2$ and keep only the terms up to first order in $1/\Lambda^2$:
\bea
\label{R_H-4}
R_H&=&\frac{(4\pi\mu^2)^{\varepsilon}}{\Gamma(2-\varepsilon)}\int_{0}^{\Omega^2} dk^2(k^2)^{-\varepsilon}\bigg\{\frac{1}{2k^4}\left(-\beta^2-2\alpha\delta-2\beta\delta-\delta^2+2\delta k^2\right)+\frac{1}{\Lambda^2 k^2}\big(\beta^2\nonumber\\
&+&2\alpha\delta+\beta\delta\big)+\cdots\bigg\} \ .
\eea
Thus, it follows trivially that
\bea
\label{R_L+R_H}
R_L+R_H&=&\frac{(4\pi\mu^2)^{\varepsilon}}{\Gamma(2-\varepsilon)}\int_{0}^{\infty} dk^2(k^2)^{-\varepsilon}\bigg\{\frac{1}{2k^4}\left(-\beta^2-2\alpha\delta-2\beta\delta-\delta^2+2\delta k^2\right)+\frac{1}{\Lambda^2 k^2}\big(\beta^2\nonumber\\
&+&2\alpha\delta+\beta\delta\big)+\cdots\bigg\} \ .
\eea
All these integrals vanish within the dimensional regularization scheme. Therefore, $R_L+R_H=0$.

Finally, since $I_L+I_H=I_L\big|_{\Omega^2\rightarrow\infty}+I_H\big|_{\Omega^2\rightarrow 0}$, it follows that we can substitute (\ref{LI_L-4}) and (\ref{LI_H-4}) into (\ref{split4}) to obtain
\bea
\label{series4}
K^{(1)}_{IV}(\bar\Phi,\Phi)&\approx&\frac{\delta}{32\pi^2\varepsilon}-\frac{1}{32\pi^2}\bigg\{(\alpha+\beta+\delta)\ln\left(\frac{\alpha+\beta+\delta}{e^{1-\gamma}\Lambda^2}\right)+\frac{1}{\Lambda^2}\bigg[\beta\delta\ln\left(\frac{\alpha+\beta+\delta}{e^{1-\gamma}\Lambda^2}\right)\nonumber\\
&+&(\beta^2+2\alpha\delta)\ln\left(\frac{2(\alpha+\beta+\delta)}{e^{1-\gamma}\Lambda^2}\right)+3\alpha\beta\ln\left(\frac{3(\alpha+\beta+\delta)}{e^{1-\gamma}\Lambda^2}\right)+2\alpha^2\nonumber\\
&\times&\ln\left(\frac{4(\alpha+\beta+\delta)}{e^{1-\gamma}\Lambda^2}\right)\bigg]\bigg\} \ .
\eea
In \eqref{series4} we have discarded the terms whose integrals over the Grassmannian coordinate identically vanishes.
 
Note that the singularities $\varepsilon^{-1}$ are not completely cancelled in the sum of the contributions (\ref{LI_L-4}) and (\ref{LI_H-4}). From the formal viewpoint, as we already noticed, the result for the Model IV can be obtained from the result for the Model III through the mutual replacement $\alpha$ by $\delta$, but while the integral from $\alpha$ over the Grassmannian space vanishes, the integral from $\delta$ does not vanish. Thus, in contrast with the previous models, the K\"{a}hlerian effective potential for the Model IV is ultraviolet divergent, as we already mentioned above. Additionally, in order to remove the divergence, we can add to the theory a similar counterterm as the one used in the standard Wess-Zumino model.  Actually, in this case we have the wave function renormalization since the only divergence is a correction to the usual kinetic term $\Phi\bar{\Phi}$, just as in the Wess-Zumino model.

\section{Conclusion}
\label{Conc}
In this paper we have investigated nonlocal supersymmetric theories with generic potential and studied the effect of the non-locality on the one-loop effective potential. Four different models have been considered. Except for the first model, which admits an essentially distinct motivation, all the other reduce themselves to the local Wess-Zumino model for the case where $\Lambda\to\infty$. 

Specifically for the first model, whose one-loop effective potential is given by Eq. \eqref{series1}, we can observe a strong non-locality. In fact the complete expression only converge for $\xi<\Lambda$. Moreover, in \eqref{series_1} we have presented the first three terms of the series.

As to the Models I, II and III, we have calculated the expressions for the corresponding one-loop effective potential. They are given by  \eqref{series2}, \eqref{series3} and \eqref{series4}, respectively. As we can see the firsts terms that appear in the corresponding expansions, coincide with the one-loop Wess-Zumino effective potential \cite{AP}. Another point that we want to mention is that the leading terms of the corrections associated with the non-locality are of the order $O(\ln\Lambda)$. We note that the behavior of quantum corrections for the large $\Lambda$ in our theory, that is, the presence of logarithmically growing correction together with the $\frac{1}{\Lambda}$ suppressed corrections, is rather similar to the higher-derivative superfield theories \cite{CMP}. Actually, this our study is a generalization of \cite{CMP} for the infinite-order derivatives.

The one-loop correction to the Model IV, presents a divergent behavior. This can be seen by the first term in \eqref{series4}, which is proportional to ${\lambda^2\bar\Phi\Phi}/\epsilon$, being $\epsilon=2-\frac d2\to 0$. In fact it can be eliminated by the wave function renormalization, providing a finite effective potential.  

 A possible continuation of this study could consist in application of the nonlocality to more sophisticated supersymmetric field theories, especially to supergauge theories and supergravity. We expect to carry out these studies in our further papers.

{\bf Acknowledgements.} This work was partially supported by Conselho
Nacional de Desenvolvimento Cient\'{\i}fico e Tecnol\'{o}gico (CNPq) and FAPESP. A. Yu. P. has been partially supported by the CNPq. through the project No. 303783/2015-0, and E. R. Bezerra de Mello through the research project No. 313137/2014-5.


\begin{thebibliography}{99}
\bibitem{Grisaru96} M. T. Grisaru, M. Rocek and R. von Unge, Phys. Lett. B383, 415 (1996).
\bibitem{Modesto} L.~Modesto, Phys.\ Rev.\ D {\bf 86}, 044005 (2012).
\bibitem{Biswas} T.~Biswas, E.~Gerwick, T.~Koivisto and A.~Mazumdar,
  Phys. Rev. Lett.\ {\bf 108}, 031101 (2012).
  \bibitem{Briscese15} F. Briscese, E. R. Bezerra de Mello, A. Yu. Petrov and V. B. Bezerra, Phys. Rev. D92, 104026 (2015).
\bibitem{BK0} I. L. Buchbinder, S. M. Kuzenko, Ideas and Methods of Supersymmetry and Supergravity, IOP Publishing, Bristol and Philadelphia, 1995.
\bibitem{SGRS} S. J. Gates, M. T. Grisaru, M. Rocek, W. Siegel. Superspace or One Thousand and One Lectures in
  Supersymmetry, Benjamin/Cummings, 1983.
\bibitem{LWS} T. Lee and G. Wick, Phys. Rev. D2, 1033 (1970); K. Stelle, Phys. Rev. D16, 953 (1977).
\bibitem{Ghost} M. Ostrogradsky, Mem. Ac. St. Petersbourg VI 4, 385 (1850); R. P. Woodard, arXiv:1506.02210. 
\bibitem{NLSUSY} T. Kimura, A. Mazumdar, T. Noumi and M. Yamaguchi, arXiv:1608.01652.
\bibitem{Vaughn} M. T. Vaughn, Introduction to Mathematical Physics, Wiley-VCH, Weinheim, 2007.
\bibitem{Efimov} G. V. Efimov, Commun. Math. Phys. 5, 42 (1967).
\bibitem{BMS} T. Biswas, A. Mazumdar, and W. Siegel, JCAP 0603, 009 (2006).
\bibitem{BFOW} L. Brekke, P. G. O. Freund, M. Olson, and E. Witten, Nucl. Phys. B302, 365 (1988).
\bibitem{Trodden} M. Fontanini, M. Trodden, Phys. Rev. D83, 103518 (2011).
\bibitem{BS} M. Beneke, V. A. Smirnov, Nucl. Phys. B522, 321 (1998).
\bibitem{Smirnov} V. A. Smirnov, Applied Asymptotic Expansions in Momenta and Masses, Springer, Heidelberg, 2002.
\bibitem{BBF} T. Becher, A. Broggio, and A. Ferroglia, Introduction to Soft-Collinear Effective Theory, Springer, Heidelberg, 2015.
\bibitem{BKY} I. L. Buchbinder, S. M. Kuzenko, J. V. Yarevskaya, Nucl. Phys. B411, 665 (1994); I. L. Buchbinder, S. M. Kuzenko, J. V. Yarevskaya, Phys. At. Nucl. 56, 680 (1993).
\bibitem{GGNPS} F. S. Gama, M. Gomes, J. R. Nascimento, A. Y. Petrov, A. J. da Silva, Phys. Lett. B733, 247 (2014).
\bibitem{AP}  I. L. Buchbinder, S. M. Kuzenko, A.Yu. Petrov, Zh.V. Yarevskaya, {\it Superfield effective potential},  hep-th/9501047.
\bibitem{CMP} M. Cvetic, T. Mariz, A. Yu. Petrov, Phys. Rev. D92, 085041 (2015). 

\end{thebibliography}
\end{document}